\documentclass[twocolumn]{book}
\usepackage[pdftex]{graphicx}  
\usepackage{makeidx,universe}
\usepackage{comment}

\newcommand{\ber}{\ensuremath{^7}Be}
\newcommand{\bor}{\ensuremath{^8}B}

\newcommand{\che}{Cherenkov}

\makeauthorindex

\BookTitle{Proceedings of the XXIX PHYSICS IN COLLISION}

\CopyRight{\copyright 2009 by Universal Academy Press, Inc.}

\begin{document} 

\pagenumbering{arabic}

\chapter{ %
{\LARGE \sf
Solar neutrinos }         \\
{ \normalsize \bf          
Marco Pallavicini }      \\
{\small \it \vspace{-.5\baselineskip}
Universit\`a di Genova -  Dipartimento di Fisica \\ and INFN Sezione di Genova, \\
via Dodecaneso 33, I-16146 Genova, Italy \\
}  }




\baselineskip=10pt    
\parindent=10pt          

\section*{Abstract}      

The Sun is a powerful neutrino source that can be used to study the physical 
properties of neutrinos and, at the same time, neutrinos are a unique tool to 
probe the interior of the Sun. 

For these reasons, solar neutrino physics is both fundamental neutrino 
and solar physics. In this paper we summarize shortly the main results of the last 
three decades and then focus on the new results produced by running experiments. 
We also give a short look at already funded or proposed new projects
and at their scientific perspectives.

\section{Introduction}   
Solar neutrinos have been for long a formidable tool to study both the fundamental properties of 
neutrinos and to probe the interior of the Sun core. 

Historically, the main interest of the first generation experiments (Homestake in particular, 
\cite{bib:homestake}) was the experimental verification of the Standard Solar Model (SSM) developed 
by J. Bahcall and co-workers, and particularly the search of a definite proof that the Sun is fueled 
by the pp fusion chain.

Since the early 70's, a clear discrepancy between theory and experiment was found by Homestake 
and later confirmed with different techniques and different energies by Kamiokande \cite{bib:sk}, 
Gallex/GNO \cite{bib:gallex} \cite{bib:gno} and Sage \cite{bib:sage}.
This discrepancy used to be known as the solar neutrino problem (SNP).

It was early understood that a possible explanation (among others) of the SNP was the
existence of neutrino flavour oscillations, which may occur if neutrinos have a non-zero mass
and if the family lepton numbers are violated (global lepton number may still be conserved).

In the simplest case, neutrino oscillate through a mixing matrix that is the equivalent of the
CKM matrix in the quark sector, although other scenarios that include additional sterile
neutrinos are possible as well. Oscillations reduce the observed counting rate either because
the experiment can see only electron neutrinos (this is the case of radiochemical experiments
like Homestake, Gallex/GNO and Sage) or because the elastic cross section on electrons is lower for muon and
tau neutrinos (Kamiokande, Borexino).

\begin{figure}[t]
\begin{center}
\includegraphics[width=.48\textwidth]{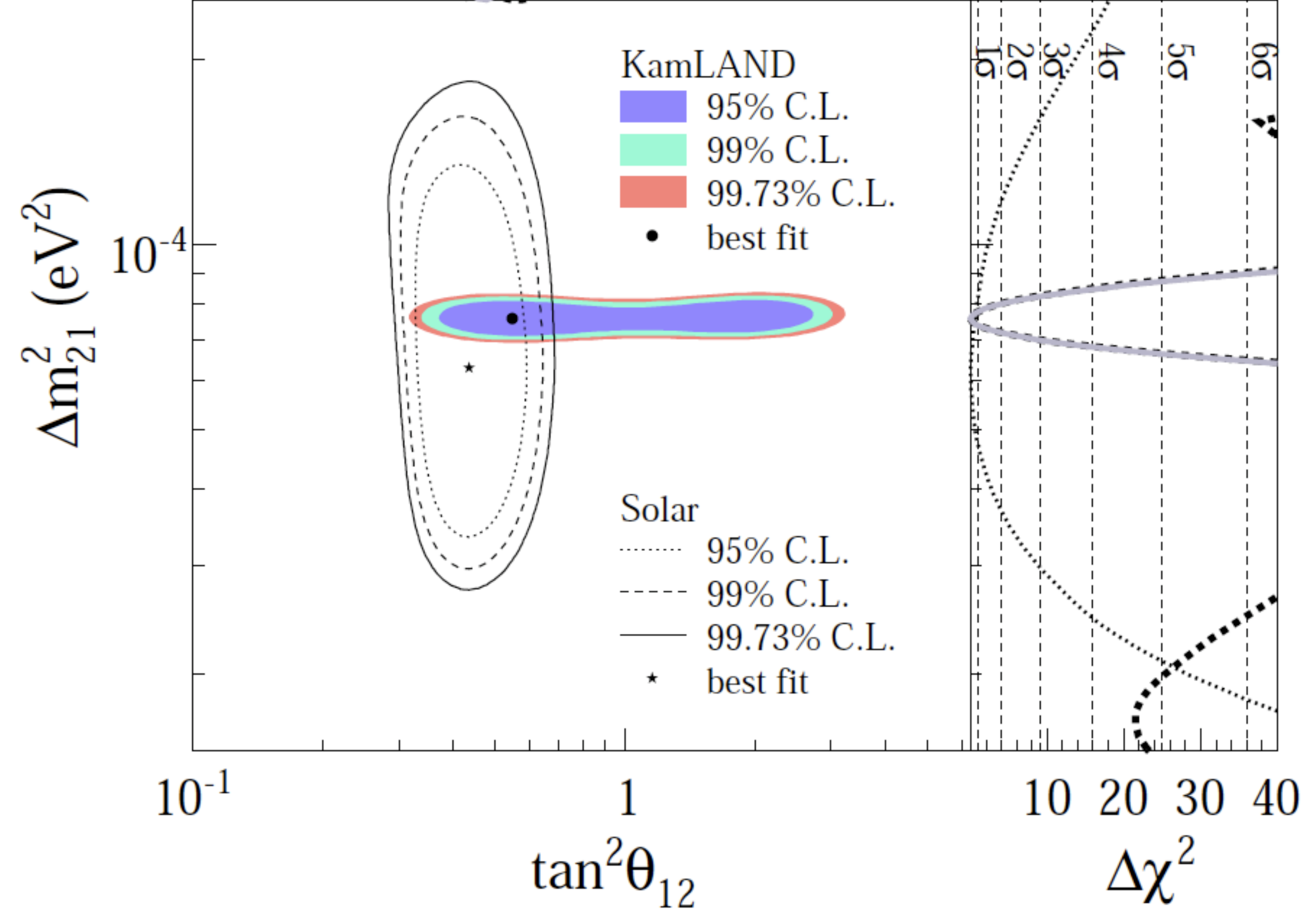}
\caption{Current knowledge of solar oscillation parameters $\Delta m^2_{12}$ and tan$^2\theta_{12}$ 
as given in Ref. \cite{bib:kamland}. The result is obtained by combining solar neutrino results and reactor data.}
\end{center}
\label{fig:parameters}
\end{figure}

A convincing proof that the oscillations are indeed the right explanation of the SNP
came in year 2001, when the SNO experiment \cite{bib:sno} proved that the total neutrino flux 
is indeed in fair agreement with SSM, while the electron neutrino component is depleted by flavor 
oscillations.
SNO could obtain this beautiful result thanks to the use of heavy water (D$_2$O) as a target. Carefully 
designed neutron detectors allowed to disentangle charged current interactions (possible for 
electron neutrinos only at the solar neutrino energies of a few MeV) and neutral current interactions 
(possible for all neutrino types), allowing the independent measurement of the electron neutrino flux 
and the total neutrino flux. The latter agrees with the SSM while the former is depleted. This is
a clear proof of neutrino oscillations. Besides, with the use of the very precise elasting scattering measurement 
given by SuperKamiokande, SNO has been able to provide the first precise measurement 
of the oscillation parameters. This measurement was later confirmed and refined by
the KamLand experiment \cite{bib:kamland}, that could observe anti-neutrino oscillation in disappearence mode from nuclear reactors
and constraint more precisely the $\Delta m^2$. The best result so far obtained for the
oscillation parameters is shown in Fig. \ref{fig:parameters}

With the conclusion of SNO, the first generation of solar neutrino experiments was over
and the existence of neutrino oscillations was established. However, several important issues
related to neutrino physics and solar physics still require an answer and a new generation of 
experiment, either brand new or upgrading existing one, is in progress.

The main goal that is related to neutrino physics is the precise experimental test
of the electron neutrinino survival probability that is computed including neutrino oscillations
in vacuum and in matter. The current solution of the SNP is based on the so called
LMA-MSW model, where matter effects in the Sun play a crucial role in explaining why
the neutrino deficit depends on neutrino energy. Although the model is in fair agreement with the existing data,
the only precise measurement are the one done by Superkamiokande and SNO, all
obtained so far with water \che\ detectors with energy threshold of 5 MeV or more (see Fig. 2). 
Low energy data before Borexino are limited to those provided by radiochemical experiments, which lack 
precision and integrate the flux in the whole energy range above detection threshold.

A precise test of the $P_{ee}$ as a function of energy is mostly important to validate the 
current scenario and search for new physics. To do so, more measurements of the solar
neutrino spectrum, and particularly those in the vacuum region (below 1 MeV) and in the
transition region (1-3 MeV) are crucial.

As already mentioned, besides a precise verification of the expected $P_{ee}$, precise
solar neutrino measurement could be sensitive to new physics, like anomalous neutrino
magnetic moments, non-standard interactions, or even probe $\theta_{13}$. 

Precision measurement of solar neutrino fluxes are very important also for solar physics,
particularly to understand what is the metallicity content of the core and what is the
role of the CNO cycle in the Sun. In main sequence stars like the Sun the CNO contribution
to the energy production is expected to be of the order of 1\%, but the precise value is
unknown. Besides, the measurement of the metal content in the core is of course not 
possible, and the models that are required to infer this value from metallicity mesurements
in the solar corona are difficult, and their result is still debated. This is the so called
metallicity controversy. Old calculations done with 1D numerical models were in fair 
agreement both heliosimology measurements (very accurate) and 
with the available neutrino data (rather poor on this matter). New improved calculations
with 3D models have spoiled this nice agreement and the situation is now unclear. 
A precise measurement of \ber\ flux could help to clarify this matter, as it is shown 
in Table \ref{tab:cno}  A direct measurement of the CNO neutrino flux, even with
moderate precision, would be decisive to clarify the role of CNO cycle in the Sun \cite{bib:carlos}.

At the time of writing, the running solar neutrino experiments are Borexino, Sage, Superkamiokande
and KamLand. The SNO collaboration is upgrading the detector to start a new solar
neutrino phase with liquid scintillator instead of heavy water (SNO+). 
Other experiments that are under discussion include solar neutrino programs. 
In the following we review the main results of the running experiments and 
outline the future perspectives of some of the others.

\begin{table}[htdp]
\begin{center}
\begin{tabular}{cccc} \hline \hline
source & BPS08 (GS) & BPS08 (AGS) & Difference (\%) \\ \hline
pp        &  5.97$\pm$0.006 &  6.04$\pm$0.005 & 1.2\% \\
pep      &  1.41$\pm$0.011 &  1.45$\pm$0.010 & 2.8\% \\
hep      &  7.9$\pm$0.15 &  8.22$\pm$0.15 & 4.1 \% \\
\ber\      &  5.07$\pm$0.06 &  4.55$\pm$0.06 & 10\% \\
\bor\       &  5.94$\pm$0.11 &  4.72$\pm$0.11 & 21\% \\
$^{13}$N    &  2.88$\pm$0.15 &  1.89$\pm$0.14 & 34 \%\\
$^{15}$O       &  2.15$\pm$0.17 &  1.34$\pm$0.16 & 31\% \\
$^{17}$F        &  5.82$\pm$0.16 & 3.25$\pm$0.15  & 44\% \\ \hline \hline

\end{tabular}
\end{center}
\caption{Expected neutrino flux as in ref. \cite{bib:carlos} assuming a high metallicty content (old, GS model)
or the lower value indicated by improved 3D calculations (new, AGS model). Large differences between the two models
are clear in \ber\ , \bor\ , and CNO fluxes. A measurement of \ber\ with 5\% precision
or better would be very useful to clarify the metallicity controversy. A meaurement of CNO neutrinos would
be probably conclusive. The table presents the predicted fluxes, in units of
10$^{10}$(pp), 10$^9$( \ber\ ), 10$^8$(pep, $^{13}$N,$^{15}$O), 10$^6$( \bor\ ,$^{17}F$), and
10$^3$ (hep) cm$^{?2}$ s$^{?1}$.}
\label{tab:cno}
\end{table}%

\begin{figure}[t]
\begin{center}
\includegraphics[width=.48\textwidth]{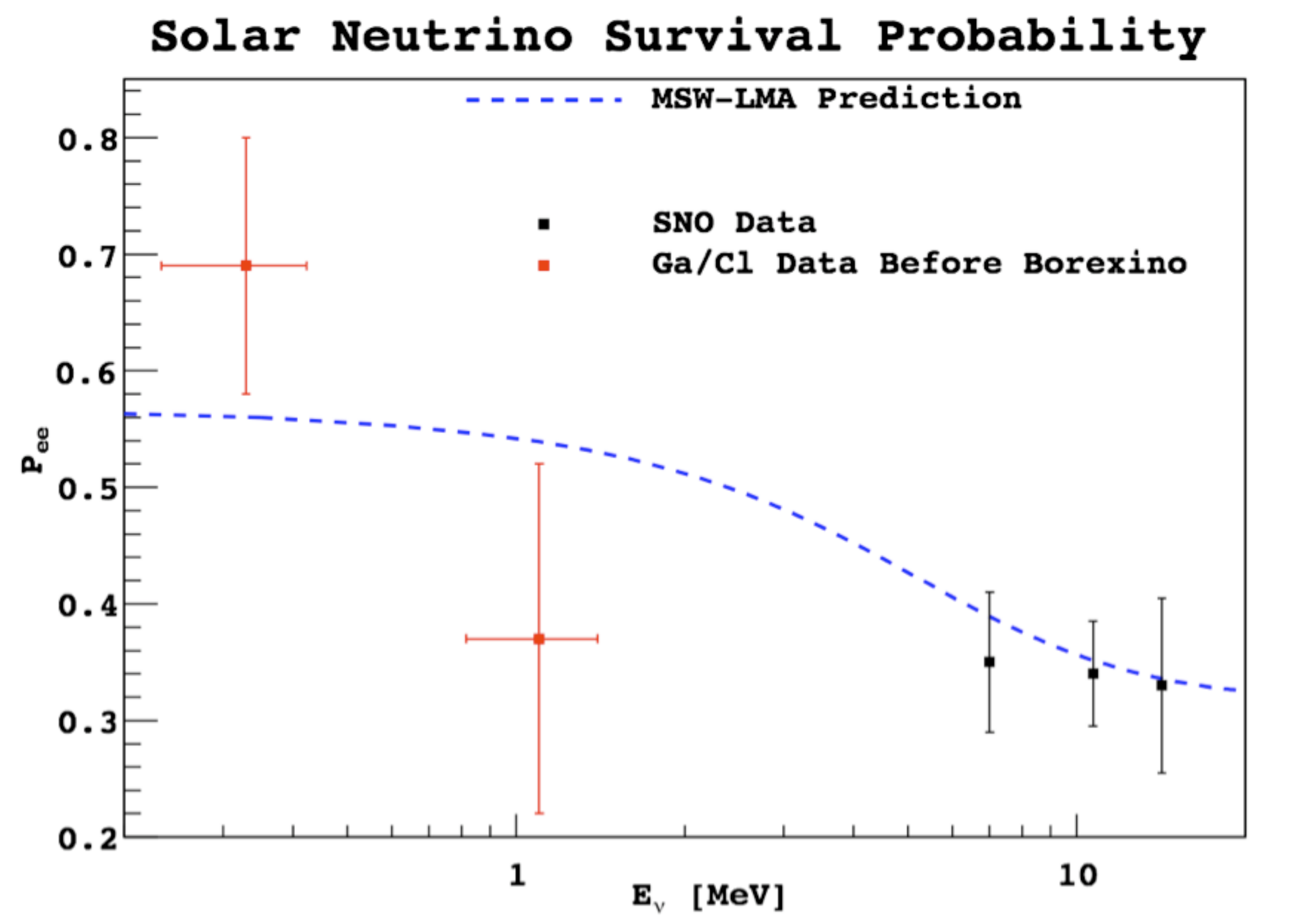}
\caption{The electron neutrino survival probability P$_{ee}$ predicted by the LMA-MSW solution 
compared to available data before Borexino. Precision
measurement were available only above 5 MeV.}
\end{center}
\label{fig:pee1}
\end{figure}

\begin{figure}[t]
\begin{center}
\includegraphics[width=.48\textwidth]{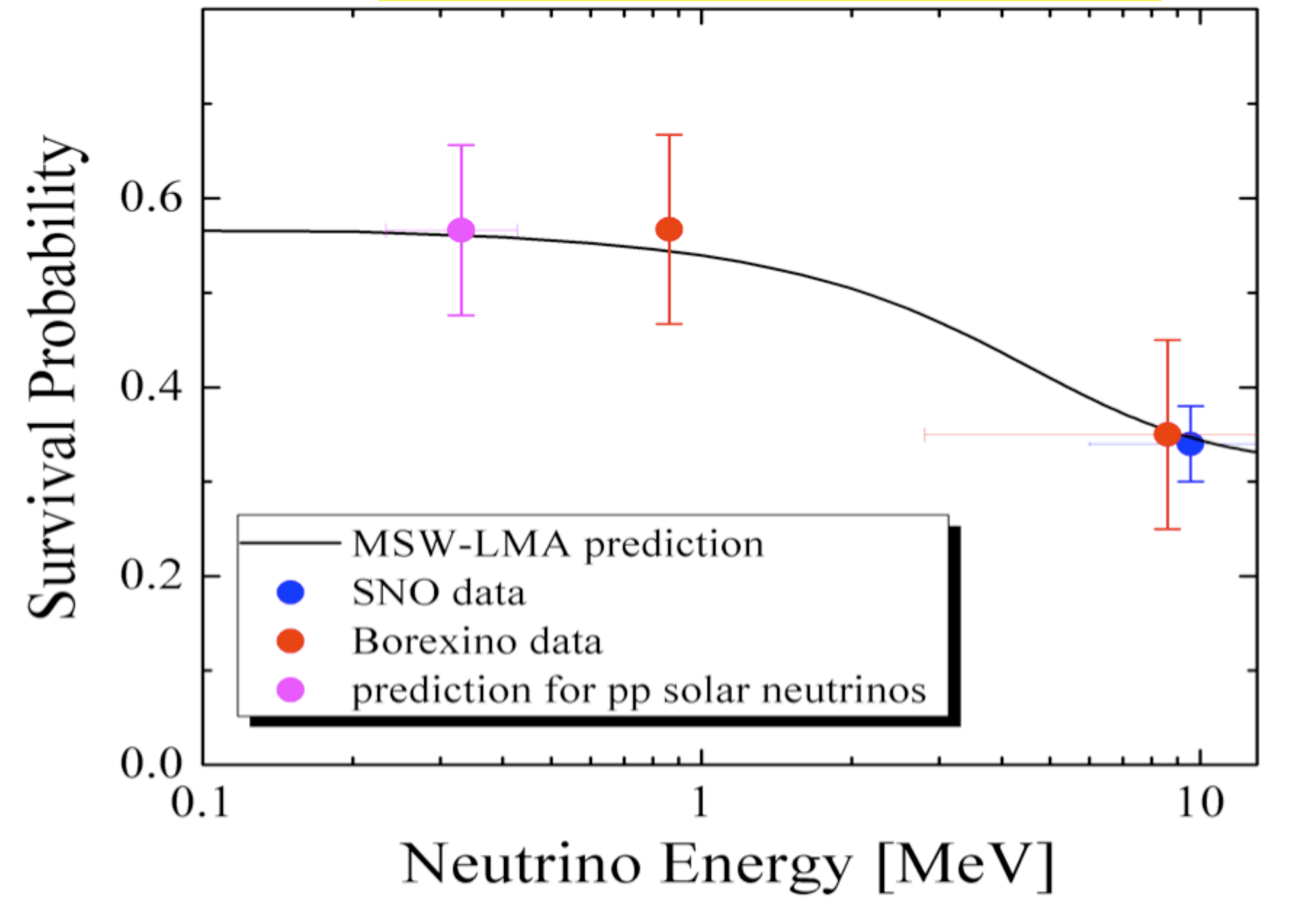}
\caption{The same curve as in Fig. \ref{fig:pee1} after first Borexino results. The flux of
$^7$Be neutrinos is now known at 12\% level, and a precision determination of pp flux is possible by means
of luminosity constraint to solar model. }
\end{center}
\label{fig:pee2}
\end{figure}

\section{Borexino}

The main solar neutrino experiment that is running at the time of writing is Borexino \cite{bib:Borex1,bib:Borex2,bib:Borex3},
a large volume liquid scintillator detector whose primary 
purpose is the real-time measurement of low energy solar neutrinos. 
It is located deep underground ($\simeq$ 3800~ meters of water equivalent, m w.e.) in the Hall C of the 
Laboratori Nazionali del Gran Sasso (Italy).

The original main goal of the experiment was the detection of the monochromatic neutrinos 
that are emitted in the electron capture decay of \ber\ in the Sun \cite{bib:Borex1}. 
However, the observed radioactive background is much lower than expected, 
which results in a potential broadening of the scientific scope of the experiment. 
Particularly, Borexino now also aims at the spectral study of other solar neutrino components, such 
as low energy $^{8}$B neutrinos, and possibly pep, pp and CNO neutrinos \cite{bib:c11}.
Besides solar physics, the unprecedented characteristics of its apparatus 
make Borexino very competitive in the detection of anti-neutrinos 
($\bar{\nu}$),  particularly those of geophysical origin. The physics goals 
of the experiment also include the detection of a nearby supernova, the 
measurement of the neutrino magnetic moment by means of a powerful neutrino 
source, and the search for very rare events like the electron decay 
\cite{bib:edecay} or the nucleon decay into invisible channels 
\cite{bib:nucleon}.

\begin{figure}[t]
\begin{center}
\includegraphics[width=.48\textwidth]{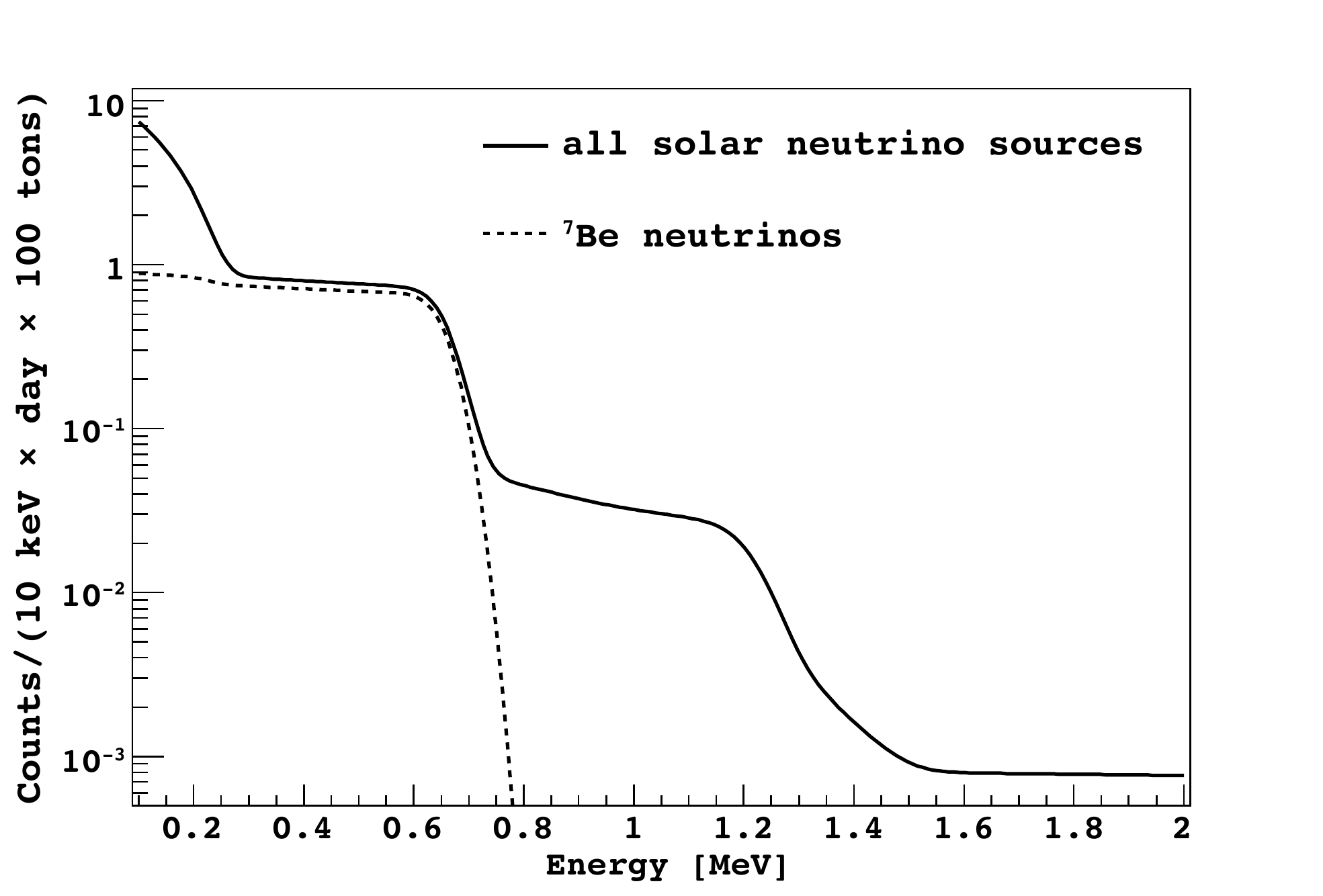}
\caption{The expected solar neutrino spectrum in a liquid scintillator with 100 t of active target. The two
shoulders around 0.7 MeV and 1.3 MeV are due to monochromatic \ber\ and pep neutrinos and 
are the main features that can be used to extract these signals. The \ber\ component is also shown separately
as dotted curve. }
\end{center}
\label{fig:NeuSpectrum}
\end{figure}

In Borexino low energy neutrinos ($\nu$)  of all flavors
are detected by means of their elastic scattering of electrons or, in the 
case of electron anti-neutrinos, by means of their inverse beta decay on protons
or carbon nuclei. Fig. 4 show the distribution of the electron recoil
energy for all solar neutrino components in liquid scintillator, normalized to 100 t of target. 

The electron (positron) recoil energy is converted 
into scintillation light which is then collected by a set of photomultipliers. A
sketch of Borexino apparatus is shown in Fig. 5  For more 
details about the detector see \cite{bib:Borex3}.

\begin{figure}[t]
\begin{center}
\includegraphics[width=.48\textwidth]{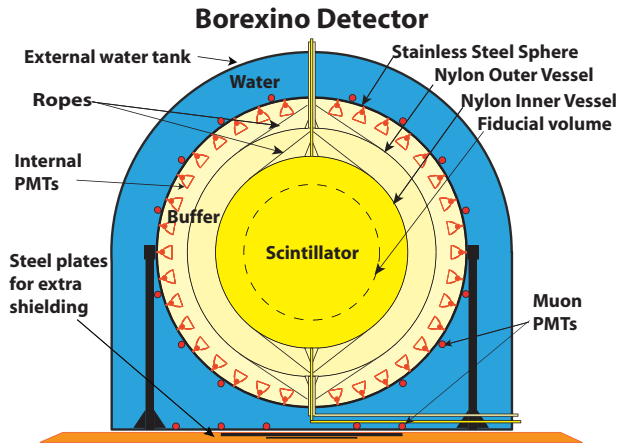}
\caption{Schematic structure of the Borexino detector.}
\end{center}
\label{fig:detec}
\end{figure}

This technique has several advantages over both the water \che\ detectors and 
the radiochemical detectors used so far in solar neutrino experiments. 
Water \che\ detectors, in fact, can not effectively detect solar neutrinos
whose energy is below several MeV, both because the \che\ light 
yield is low and because the intrinsic radioactive background cannot be 
pushed down to sufficiently low levels. On the other hand, radiochemical experiments 
cannot intrinsically perform spectral measurements and do not detect events in
real time.

An organic liquid scintillator solves the aforementioned problems: the low
energy neutrino detection is possible because of the high light yield,
that in principle allows the energy threshold to be set down to a level of a 
few tens of keV\footnote{However,
the unavoidable contamination of $^{14}$C that is present in any organic
liquid practically limits the "neutrino window" above $\approx$ 200 keV}; the
organic nature of the scintillator, and its liquid form at ambient temperature,
provide very low solubility of ions and metal impurities, and yield the technical 
possibility to purify the material as required. However, no measurement of
the direction of the incoming neutrino is possible and, even more importantly, 
the neutrino induced events are intrinsically indistinguishable from 
$\beta$ and $\gamma$ radioactivity, posing formidable requirements in
terms of radiopurity of the scintillator and of the detector materials.

\begin{figure}[h]
\begin{center}
\includegraphics[width=0.48\textwidth]{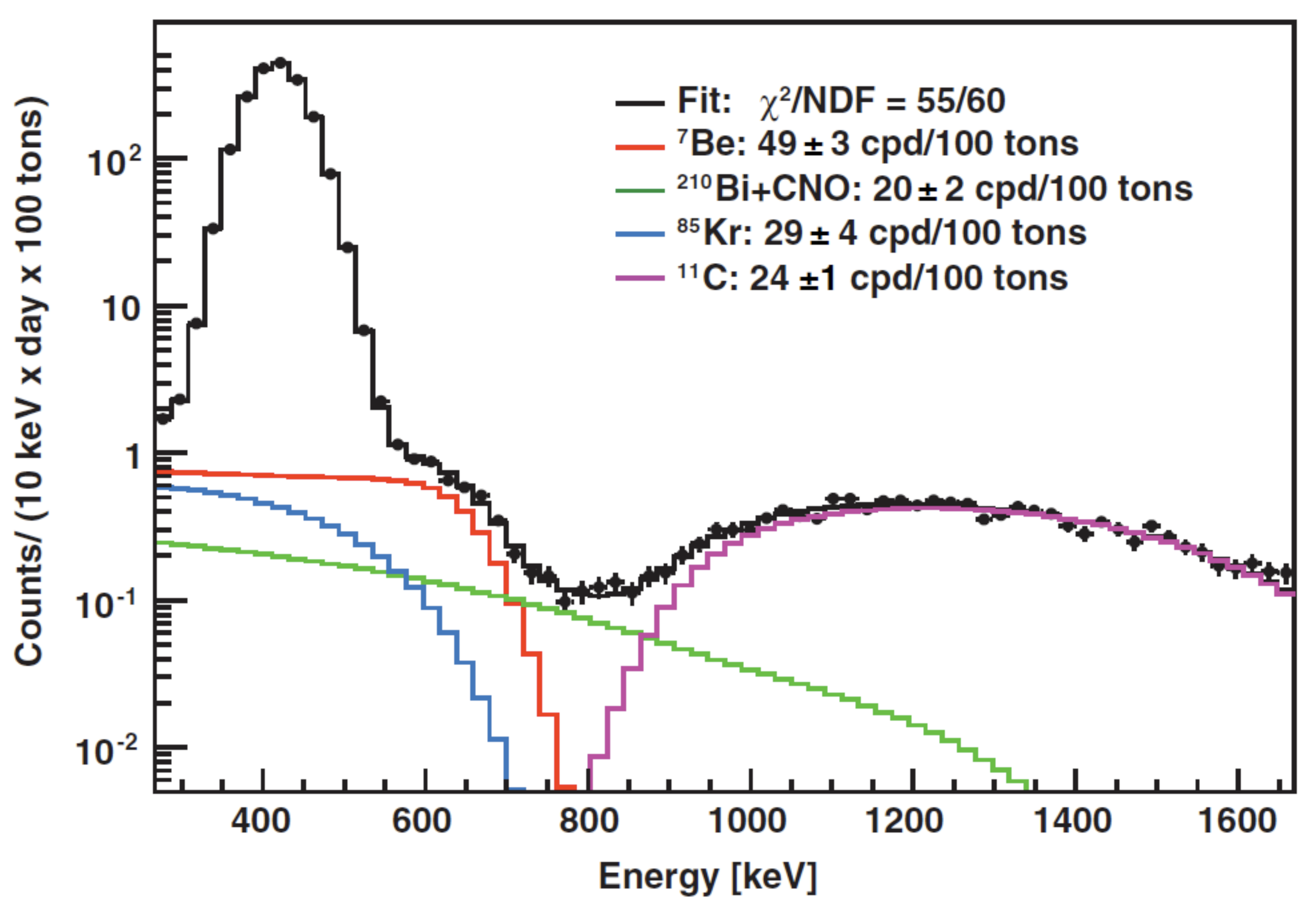}
\caption{The spectrum in the energy region between 300 keV and 1700 keV obeserved by Borexino with
192 days of live time. See \cite{bib:neutrinopaper} for details. }
\label{fig:be7}
\end{center}
\end{figure}

As shown in \cite{bib:be7paper} and \cite{bib:neutrinopaper} these requirements
have been all met, and sometimes also exceeded, yielding the first measurement
of \ber\ solar neutrino flux and of the $^{8}$B neutrinos with a electron energy
threshold of 2.8 MeV \cite{bib:boro8}. Fig. \ref{fig:be7} shows the spectrum measured
by Borexino with 192 days of live time in the energy region between 300 and 1700 
keV of electron recoil energy. The characteristic "Compton edge" of the monocromatic
\ber\ neutrinos is clearly visible in the spectrum. Fit results yield a rate 
of 49 $\pm$ 3 $\pm$ 4 cpd/100 ton where the first error is statistical and the second error
is systematic. See \cite{bib:neutrinopaper} for more details.

In the high energy region of Fig. \ref{fig:be7} a large bump between above 900 keV
is visible. This is largely due to the $\beta^+$ decay of cosmogenically produced 
$^{11}$C and is the main sorce of background for a possible future measurement of 
pep and CNO neutrinos. This background is un-avoidable in liquid scintillator detectors
and its size depends on the depth of the laboratory in which the detector is located. 
All runnning and future experiments will have to deal with it. The Borexino
collaboration has developed a technique (\cite{bib:c11}) to tag this events by exploting
the fact that the production of $^{11}$C is accompanied in 95\% of the case with at
least one emitted neutron. The triple coincidence among the neutron, the captured neutron
and the $^{11}$C decay can be used to tag and remove these events. This technique
has been already validated in the Counting Test Facility.

The current precision in the \ber\ neutrino measurement is limited by lack of energy
calibration and by the still incomplete knowledge of the detector response function.
Borexino has undertaken in 2009 several calibration campaigns with $\alpha$, 
$\beta$, $\gamma$ and neutron sources located in different positions of the 
detector volume, within and outside the neutrino fiducial volume used for flux
measurements. After a complete analysis of the calibration data, a 5\% (or better) 
measurement of \ber\ flux is anticipated. This result will be a strong check of the
LMA-MSW scenario and may give important hints on the Sun metallicity problem.

Finally, the Borexino collaboration is getting ready for a purification campaign in early 2010.
The goal of this effort is to reduce $^{85}$Kr content in the scintillator (the main source
of statistical error in the \ber\ region and a severe background for a possible future
measurement of pp neutrinos) and to reduce the $^{210}$Bi background. If this effort
will be successfull, Borexino will try to measure pep and possibly CNO neutrinos, and
may also search for a pp neutrino signal. 

\section{Superkamiokande, SNO and KamLand}
The Superkamiokande experiment is a large water \che\ detector located in the Kamioka mine, in Japan. 
The importance of this experiment in solar, atmospheric and supernova neutrino physics is so
significant that does not need to be reviewed here. 
We just focus on the future perspectives in solar physics \cite{bib:sk1} \cite{bib:sk2}  \cite{bib:sk3}.

The main current goal of the experiment in solar neutrino physics is the reduction of the energy threshold. 
So far, SK has observed solar neutrinos with an electron recoil energy threshold of about 5 MeV 
(changed slightly in the different phases of the experiment). 
In order to probe the transition region of the P$_{ee}$ a lower threhsold is necessary.
SK has recently upgraded the electronics and has installed a new purification system of the water.
With better triggering, better electronics and lower radioactive background from U and Th daughters
dissolved in water, the experiment aims to push the threhsold below 4 MeV and an analysis is in 
progress with a threshold of 4.5 MeV. If this will be successfull,
the experiment should be sensitive to the raise of the  P$_{ee}$ at lower energy, the so called "energy 
upturn". This would yield a very nice check of the LMA-MSW and probe new physics.

\begin{figure}[h]
\begin{center}
\includegraphics[width=0.48\textwidth]{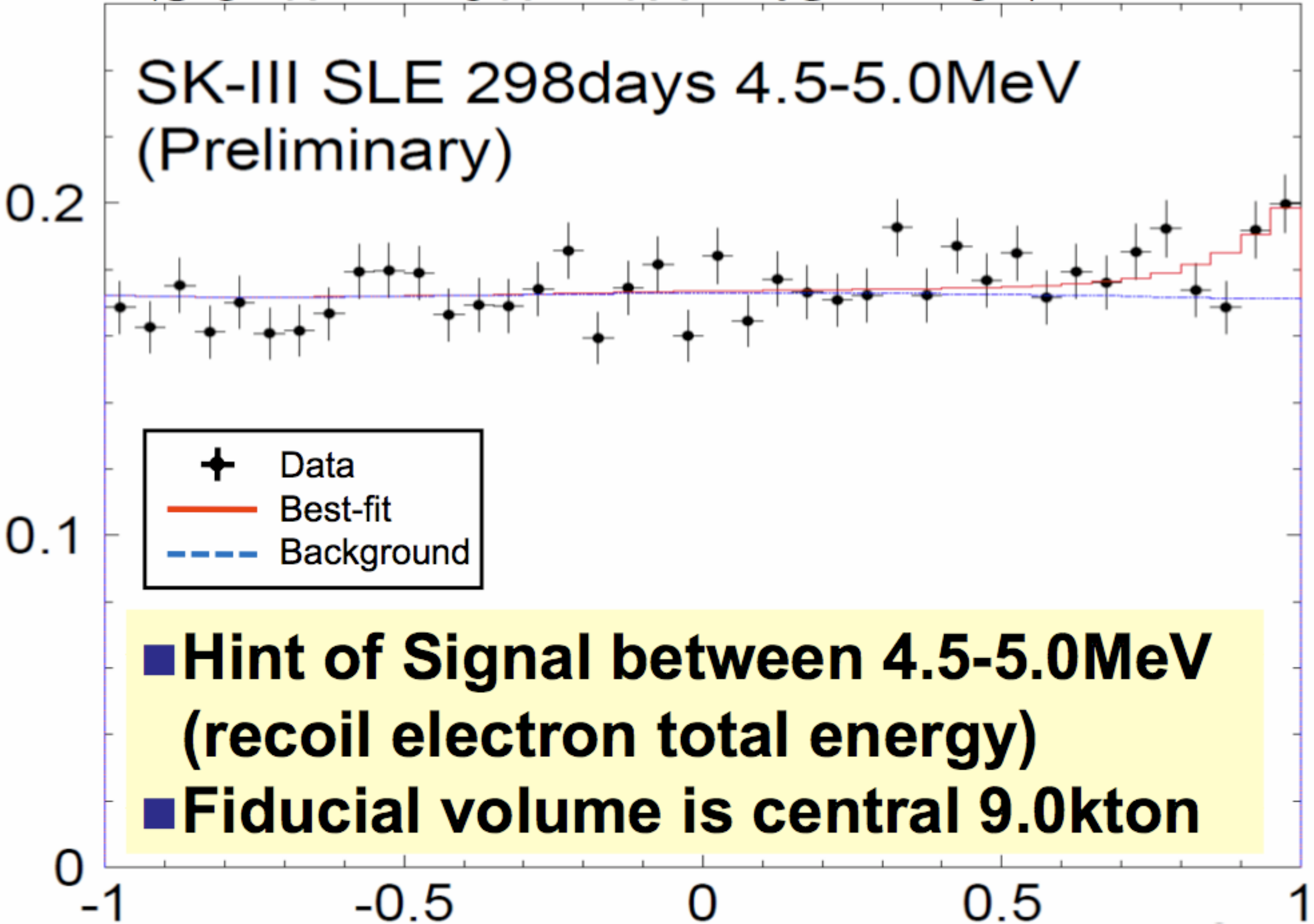}
\caption{The SK-III preliminary hint of evidence of solar neutrino signal for events in the energy
region between 4.5 a 5 MeV in electron recoil energy \cite{bib:smy2009}. }
\label{fig:sk1}
\end{center}
\end{figure}

\begin{figure}[h]
\begin{center}
\includegraphics[width=0.48\textwidth]{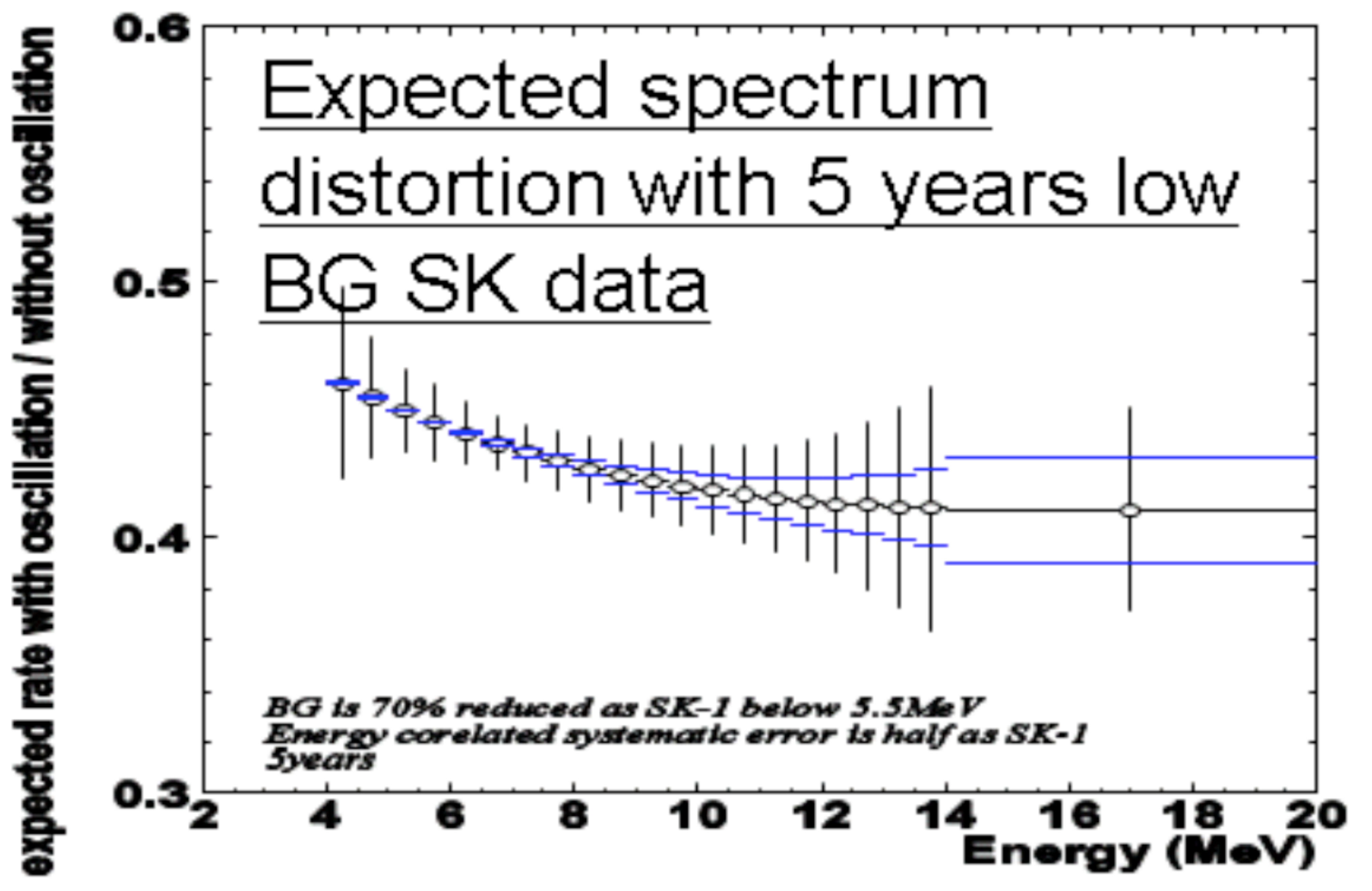}
\caption{The expected sensitivity to the energy upturn for 5 years of data taking in SK-IV with
low energy threshold and low background. }
\label{fig:sk2}
\end{center}
\end{figure}

The SNO experiment has finished data taking in 2008 after a very successfull program
where neutrino oscillations have been established. The current effort, besides the detector upgrade
program that we will cover in the next section, is to re-analyze the available data with a lower
energy threshold. A new code has been developed for this task and the goal is to study the 
neutrino signal down to 3.5 MeV. Even for SNO, the scientific reason for this is to study
the energy upturn and probe P$_{ee}$ in the transition region.

KamLand experiment has begun in 2007 a strong effort to purify the scintillator and reach the 
background levels required for solar neutrino physcis. Purification campaigns have been done
in 2007 (Apr. - Aug. ) and from June 2008 until Feb. 2009, reducing the backgrounds of 4-5 orders of magnitude. 

According to \cite{bib:kamland2}, the second purification campaign was very successfull, and after a few
months needed to get a stable background ($^{210}$Bi decay), the solar neutrino run has begun.

The collaboration plans to measure the \bor\ neutrino flux with an error of 10 \% and with 
an energy threshold of 3 MeV, and the \ber\ flux with an error of 13 \%. 
Both measurements will be completed in two years of data taking.

\section{Future projects}
Several ambitious experiments are under construction or are being considered by the scientific 
community. For lack of space it is not possible to give a comprehensive review of all proposed 
projects. I will limit the discussion for those who are already approved or very mature for approval.

The most important future project in solar neutrino physics, besides of course those already running, 
is SNO+ \cite{bib:snop}.

The SNO+ collaboration aims at the construction  of a liquid scintillator experiment re-using the 
vessel and the photomultipliers of the SNO experiment. The scintillator is linear alkylbenzene 
(LAB) doped of PPO, which is safe (high flash point), low cost and compatible with the existing acrylic vessel.
A holding net will be installed to counterbalance the buoyancy force that will develop as 
soon as the scintillator will be put in the vessel (the external shielding is still done with water, which
is denser than LAB+PPO).

\begin{figure}[h]
\begin{center}
\includegraphics[width=0.48\textwidth]{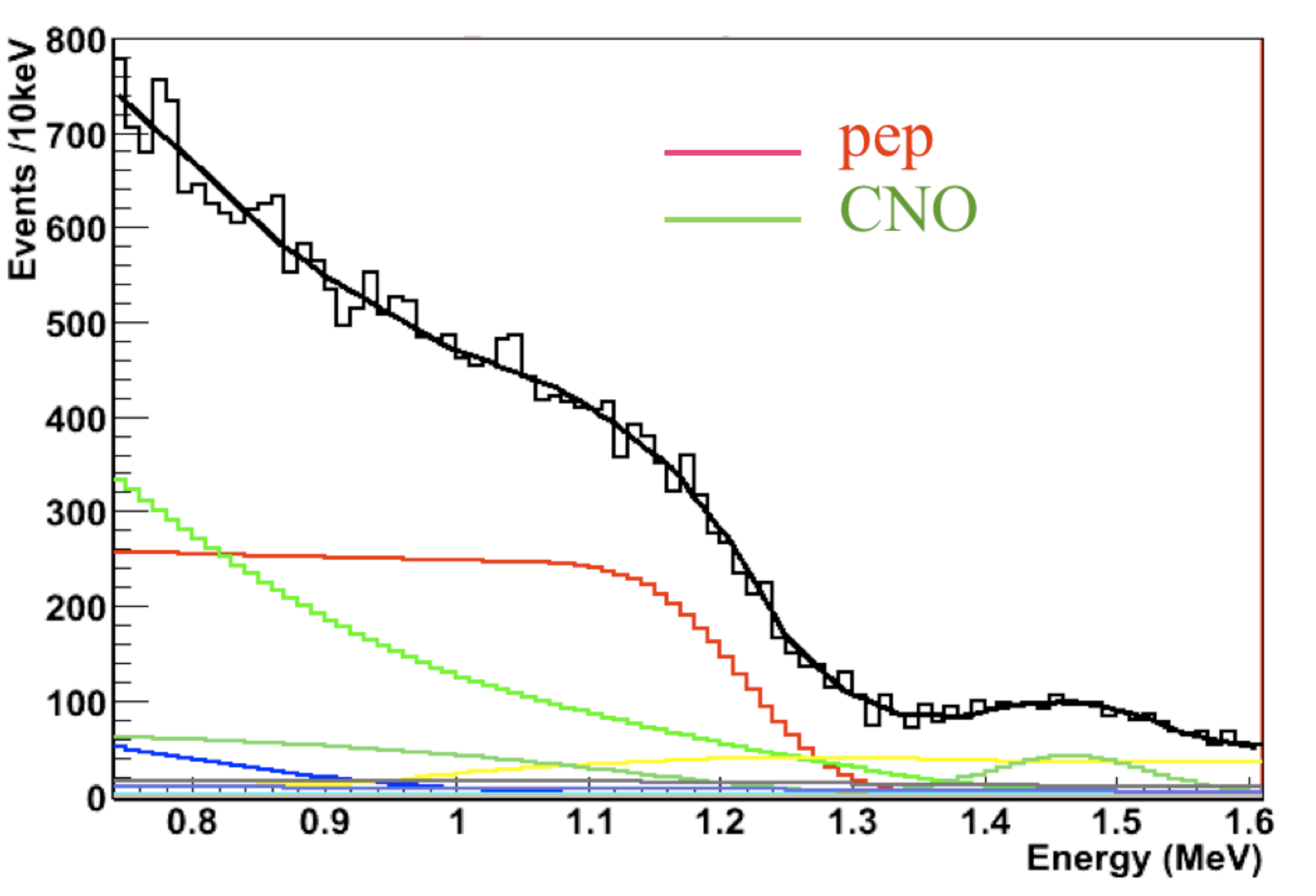}
\caption{The signal expected in SNO+ in three years of data taking for CNO and pep neutrinos. The
low cosmogenic $^{11}$C background and the high mass make SNO+ may allow a 6\% measurement
of CNO neutrinos in three years of data. }
\label{fig:sno}
\end{center}
\end{figure}

The project aims at the measurement of pep and CNO neutrinos with high precision. The higher 
depth of the Sudbury mine compared to Gran Sasso and Kamioka yields a much lower cosmogenic
background from $^{11}$C (700 times less than Kamioka, 100 times less than LNGS). 
According to the simulations, SNO+ may be able to measure
the CNO neutrino flux with 6\% error. This precision is sufficient to measure the metal content of the
solar core in a model independent way and possibly solve the solar metallicity problem.
Besides, the project aims at the measurement of pep neutrinos. These are very important to probe
P$_{ee}$ in the transition region. They are monochromatic neutrinos, and their production rate in
the Sun is very well known because its value is strongly correlated to solar luminosity. Therefore,
a precision measurement of the pep flux at Earth yields a precise measurement of P$_{ee}$  with
small uncertainties. The expected signal in SNO+ for pep and CNO neutrinos is shown in 
Fig. \ref{fig:sno}

Another more ambitious project for solar neutrino physics is LENS, Low Energy Neutrino
Spectroscopy. This experiment aims at the precise measurement of all solar neutrino components
by detecting neutrinos via charged current interactions (inverse beta decay) with $^{115}$In. 

The experimental tool used in the LENS detector
for the detection of solar neutrinos is the tagged
capture of $\nu_e$Õs on $^{115}$In via charged current inverse beta decay. The
tagged technique has two outstanding
advantages over competing scattering experiments:
First, there is a one-to-one correspondence between
the incoming neutrino energy and the measurable
electron energy E$_\nu$=E$_e$+Q$_d$ (Q$_d$: capture threshold),
and second, the $\gamma$-cascade allows the application of
time/space coincidence techniques to suppress
ubiquitous radioactive backgrounds as well as the
inherent background from the beta decay of $^{115}$In.
A modular detector is required. The LENS detector is a novel ÒScintillation
Lattice ChamberÓ, an optically segmented, three dimensional
array of ~0.5 l cells of liquid scintillator
loaded with ~8-10 wt\% Indium. The
scintillation signal from each cell will be always
viewed by the same set of 3-6 phototubes. Thus, the
full scale LENS of ~125 tons InLS, though large in
size, is in essence, a large array of small detectors
capable of bench-top precision nuclear
spectroscopy. It will provide extraordinary spatial
resolution in a large mass of liquid scintillator
through segmentation rather then time of flight
information, which allows adequate background
rejection using the time/space coincidence tag even
for low energy (~100 keV) events.

\begin{figure}[h]
\begin{center}
\includegraphics[width=0.48\textwidth]{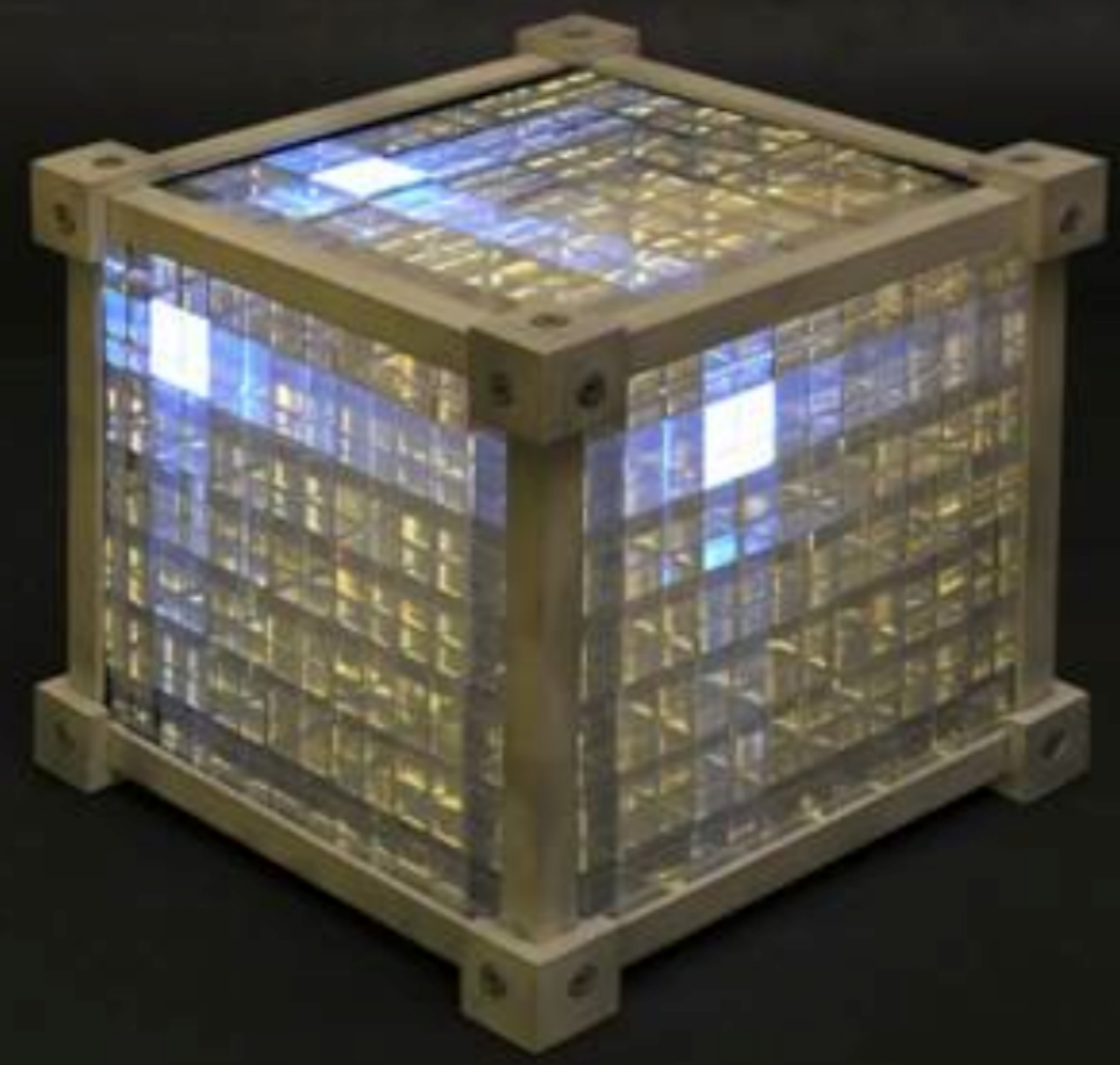}
\caption{Lens conceptual design.
Isotropically emitted scintillation light is channeled to the 
outside of the detector along the main axes via total 
internal reflection, providing direct information about the 
location of the event. }
\label{fig:lens1}
\end{center}
\end{figure}

\begin{figure}[h]
\begin{center}
\includegraphics[width=0.48\textwidth]{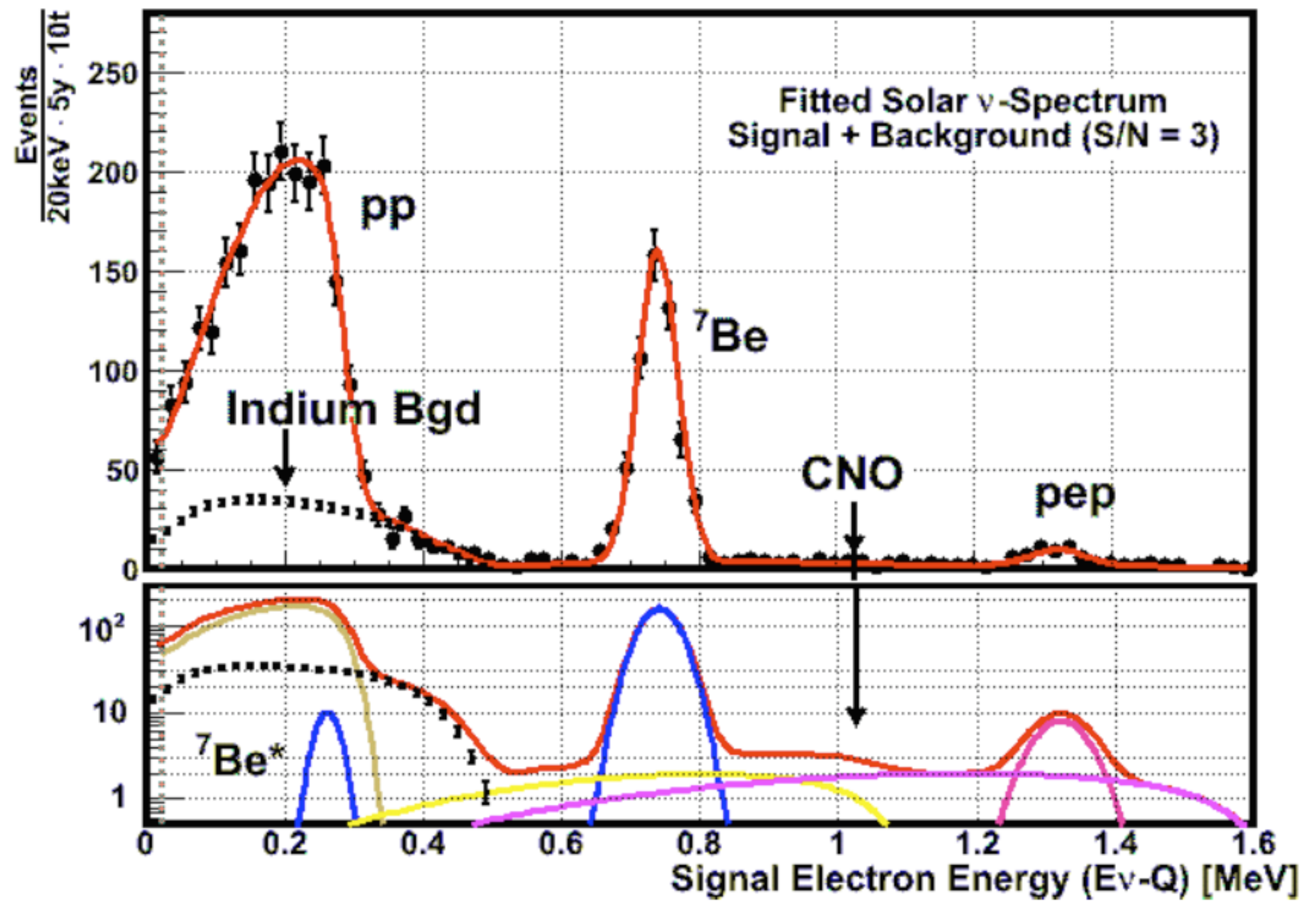}
\caption{Lens expected spectral resolution on solar neutrinos. If the detector will work as anticipated, 
LENS will give a beautiful spectral measurement of all solar neutrino components separately, with
very high precision and low background. }
\label{fig:lens2}
\end{center}
\end{figure}

The LENS collaboration is
funded by NSF for a complete R\&D of a detector based on Indium immersed in liquid scintillator modules,
as shown in Fig. \ref{fig:lens1}  If successfull, this experiment might really represent the future of solar
neutrino physics because of the superbe energy resolution possible for all neutrino energies. A spectrum
of the expected signal in LENS is shown in Fig. \ref{fig:lens2}

Finally, several new projects include solar neutrino in their scientific program. Among those, we briefly mention
Clean \cite{bib:clean} and Xmass \cite{bib:xmass}. We refer to the references for more details. All these programs want to use
the nice features of criogenic liquid as active targets for solar neutrinos and search for neutrinoless 
double beta decay and dark matter. Particularly, liquid noble gasses have a good light yield, high
density and can be purified thoroughly, yielding good signals and low bakground. Large mass detectors
of this type maybe a crucial tool for next generation solar neutrino experiments.

\section{Conclusions}
Solar neutrino physics is still a very exciting field of reasearch. After the end of the first generation of
experiments, and the establishment of neutrino flavour oscillations, several important topics
in neutrino physics and solar physics can be probed via a careful and precise measurement
of solar neutrino components. Borexino is currently running at Gran Sasso and is
giving fundamental results in low energy neutrino physics, particularly on \ber\ and \bor\ 
neutrinos. Other experiments like KamLand
and Kamiokande will soon be able to probe the low energy region as well, and test
the LMA-MSW solution. Future projects like SNO+ and possibly LENS will make another
big step forward, and open the era of high precision solar neutrino spectroscopy.
Other more ambitious projects like Clean, Deap and Xmass may join in the near future.



\end{document}